\documentclass{ecai} 

\usepackage{latexsym}
\usepackage{amssymb}
\usepackage{amsmath}
\usepackage{amsthm}
\usepackage{booktabs}
\usepackage{enumitem}
\usepackage{graphicx}
\usepackage{tabularx}
\usepackage{color}
\usepackage{tikz}
\usetikzlibrary{shapes, decorations.pathreplacing}

\newtheorem{theorem}{Theorem}

\newtheorem{corollary}[theorem]{Corollary}
\newtheorem{proposition}[theorem]{Proposition}
\newtheorem{observation}[theorem]{Observation}

\newtheorem{definition}{Definition}
\newtheorem{example}{Example}

\newcommand{\red}[1]{\textcolor{red}{\textrm{#1}}}

\usepackage{pifont}
\newcommand{\cmark}{\ding{51}}%
\newcommand{\xmark}{\ding{55}}%

\newcommand{\BibTeX}{B\kern-.05em{\sc i\kern-.025em b}\kern-.08em\TeX}

\begin{document}
\tikzset{
 vote/.style = {draw, circle, thick, fill = white, minimum size = 2.5mm, inner sep = 0pt},
 param/.style = {draw, shape = diamond, thick, fill = white, minimum size = 3mm, inner sep = 0pt},
 outcome/.style = {draw, shape = rectangle, thick, fill = white, minimum size = 2mm, inner sep = 0pt}
}

\begin{frontmatter}


\paperid{969} 


\title{Analyzing Incentives and Fairness in Ordered Weighted Average for Facility Location Games}


\author[A]{\fnms{Kento}~\snm{Yoshida}}
\author[A]{\fnms{Kei}~\snm{Kimura}\orcid{0000-0002-0560-5127}}
\author[A]{\fnms{Taiki}~\snm{Todo}\orcid{0000-0003-3467-329X}}
\author[A]{\fnms{Makoto}~\snm{Yokoo}\orcid{0000-0003-4929-396X}}


\address[A]{Graduate School of Information Science and Electrical Engineering, Kyushu University}

\begin{abstract}
 Facility location games provide an abstract model of
 mechanism design. In such games, a mechanism takes a profile of $n$ single-peaked preferences over an interval as an input and determines the location of a facility on the interval. 
 In this paper, we restrict our attention to distance-based single-peaked preferences and focus on a well-known class 
 of parameterized mechanisms called {\em ordered weighted average methods}, which is proposed by Yager~\cite{Yager:SMC:1988}
 and contains several practical implementations such as the standard average and the Olympic average.
 We comprehensively analyze their performance in terms of both incentives and fairness.
 More specifically, we provide necessary and sufficient conditions on their parameters to achieve strategy-proofness, non-obvious manipulability,
 individual fair share, and proportional fairness,
 respectively.
%
%
\end{abstract}

\end{frontmatter}

\section{Introduction}
\label{sec:intro}
\begin{table*}[tb]
 \centering
 \caption{A summary of compatibility between 
 incentive and fairness properties in OWAs. Each of two symbols, \cmark and \xmark, respectively 
 indicates compatibility and incompatibility of 
 corresponding row and column conditions.
 UN stands for unanimity. NOM-B stands for 
 one of the conditions required in NOM, corresponding to the best-case.}
 \label{tbl:summary}
 \def\arraystretch{1.75}
\begin{tabularx}{\linewidth}{X |
*{3}{>{\centering\arraybackslash}X}}
                            & ~~~ PF: $w_{j} = 1/n$ for all $j$ \newline (Theorem \ref{thm:PF})     & IFS: $w_{1} \geq 1/n$ and $w_{n} \geq 1/n$
                            \newline
                            (Theorem \ref{thm:IFS})
                            & ~~~~~~~~~UN: all OWAs \newline 
                            (Observation \ref{obs:OWA:UN})\\ 
                            \hline
  SP: $w_{j} = 1$ for some $j$  \newline 
    (Proposition \ref{prp:SP})  & \xmark & \xmark & \cmark \\ 
  \hline
  NOM: $w_{j} = 1$ for some $j$ or 
  \newline 
  $w_{1} = w_{n} = 0$  (Theorem \ref{thm:NOM}) & \xmark & \xmark & \cmark \\ 
  \hline
  NOM-B: all OWAs \newline (Proposition \ref{prp:BNOM})              & \cmark & \cmark & \cmark \\
  \hline
\end{tabularx}
\end{table*}
\renewcommand{\arraystretch}{1}

Multi-agent decision making is a fundamental problem in the field of artificial intelligence, where multiple agents interact with each other, and the society containing these agents makes
a joint decision.
Multi-agent decision making has been applied to 
various domains, such as  
federated learning, multi-agent path finding, consensus building, automated negotiations, and resource/task allocations. It has been tackled by the multi-agent research research community both theoretically and practically.

From the theoretical viewpoint, game theory has played an important role in multi-agent decision making. More specifically, mechanism design
has been attracting much attention from researchers in the field of multi-agent systems. The main purpose of mechanism design
is to develop decision making rules, which 
appropriately incentivise each agent in the society to play a desirable action.
The theory of mechanism design has been used as a mathematical foundation of multi-agent decision making in the last two decades. 

Facility location game is one of the well-studied
mechanism design problem. In a standard setting,
there are $n$ agents, each of which has a house located on the street, mathematically represented 
as a line segment $A$, and a trusted third party is now willing to build a public facility, say a library, on the street.
Assume that agents prefer the library to be placed
closely to their houses.
A mechanism takes $n$ addresses of those houses, reported by the agents, and determines where to locate the library. Various mechanisms have been proposed, including the {\em generalized median voter schemes}~\cite{moulin:PC:1980}.
Such a problem has also been widely studied
in the field of social choice theory, a subfield
of microeconomics. 

The standard facility location problem can be
seen as finding a mapping/function from $A^{n}$ to $A$,
where $A$ represents the line segment.
While taking a median seems to be theoretically
well-justified, there is another class of such functions, i.e., taking a (weighted) average.
Indeed, such problems are also known as aggregation problems, and Yager~\cite{Yager:SMC:1988} proposed
a general class of functions that take an average in a broad sense, so-called
{\em ordered weighted average (OWA) methods}.
An OWA has a list of $n$ weights, $(w_{j})_{j \in \{1, \ldots, n\}}$,
where $j$-th weight $w_{j}$ will be assigned to 
$j$-th smallest input and the OWA returns the weighted average. The set of OWAs contains the standard average, the Olympic average (which first truncates the two extreme values and takes the average of the remainings), and order statistics.

Although the OWA has been widely studied in the field of fuzzy preference aggregations,
their incentive properties have not been investigated in detail, especially from the viewpoint of mechanism design.
One of the main purposes of this paper is, therefore, to completely clarify whether they satisfy some incentive properties. Given the characterization result by Moulin~\cite{moulin:PC:1980},
we begin the analysis by focusing on strategy-proofness; a mechanism is said to be strategy-proof
if, for each agent, reporting a true address is a (weak) dominant strategy.
We also consider a weaker variant of incentive property, called non-obvious manipulability, proposed by Troyan and Morrill~\cite{TROYAN2020104970}; a mechanism is
said to be not obviously manipulable if, for each agent, reporting a true address is weakly better than any other report in both the best and the worst situations.

Besides those incentive properties,
fairness is also an important criterion for 
evaluating mechanisms~\cite{FREEMAN2021105234}.
As an extreme example, the dictatorship mechanism
that places the facility at the reported address of
a pre-determined agent is quite unfair, although 
it is strategy-proof.
In this paper we consider a series of {\em proportionality-based fairness} properties, introduced by Aziz et al.~\cite{aziz2023strategyproof}, and analyze whether OWAs satisfies these properties.
More precisely, we focus on {\em unanimous fair share} and {\em proportional fairness}; the former
requires that for each subset $S$ of agents who are located at the same address, their costs, defined by the distance from their (shared) address and the location of the facility, is upper bounded by $1 - \frac{|S|}{n}$, where the first term 1 corresponds to the length of the street. The latter extends this concept for every subset of agents, and therefore implies the former. The details of
these definitions will be given in Section~\ref{sec:model}.

Our contribution in this paper is two-fold,
which is summarized in Table~\ref{tbl:summary}.
We first clarify which OWAs do (not) satisfy strategy-proofness (SP)
and non-obvious manipulability (NOM), respectively,
presented in the left column of Table~\ref{tbl:summary}. 
We show that an OWA satisfies strategy-proofness if and only if
there exists a single weight, say $w_{j}$, which is set to one.
In other words, an OWA is strategy-proof if and only if
it is represented as an order statistic.
On the other hand, an OWA satisfies non-obvious manipulability 
if and only if it is either an order statistics or 
$w_{1} = w_{n} = 0$ holds. This implies that although the standard
average is obviously manipulable,
the Olympic average, which resembles the standard average,
is not obviously manipulable.

We then clarify which OWAs do (not) satisfy proportional fairness (PF) and individual fair share (IFS), respectively, which is summarized in the top row of Table~\ref{tbl:summary}.
We show that an OWA satisfies PF
if and only if all the weights are equal to $1/n$.
In other words, the standard average mechanism is the
only OWA mechanism satisfying PF.
On the other hand, an OWA satisfies IFS
if and only if both $w_{1} \geq 1/n$ and $w_{n} \geq 1/n$ hold.
Choosing the center between the minimum and the
maximum reported addresses, which is known to be optimal for the 
maximum cost objective, satisfies IFS,
while choosing the median among all the reported addresses,
which is 
optimal for the social cost objective,
is proved to violate IFS~\cite{aziz2023strategyproof}.

This paper is organized as follows.
Section~\ref{sec:literature} reviews related works on the ordered weighted average methods, facility location games,
non-obvious manipulability analysis,
and fair facility locations.
Section~\ref{sec:model} defines the mathematical model of the paper.
Section~\ref{sec:SP} and
Section~\ref{sec:NOM}
give the necessary and sufficient condition
for OWAs to satisfy strategy-proofness and
non-obvious manipulability, respectively.
Section~\ref{sec:fairness}
then gives the necessary and sufficient conditions for OWA to satisfy proportional fairness
and individual fair share, respectively.
Finally, Section~\ref{sec:conclu} 
discusses the compatibility of agents' incentives and
fairness in OWA and raises some open questions.

\section{Literature Review}
\label{sec:literature}

Ordered Weighted Average (OWA) methods are well-known 
preference aggregation schemes, which was formally described by Yager~\cite{Yager:SMC:1988}. Since then, various discussions on OWA have been made, even in the field of artificial intellgence~\cite{DBLP:conf/aldt/ElkindI15,DBLP:conf/atal/AmanatidisBLMR15,DBLP:journals/ai/SkowronFL16}. 
Garcia-Lapresta and Llamazares~\cite{GARCIALAPRESTA2001463}
and 
Llamazares~\cite{DBLP:journals/isci/Llamazares07} studied
OWAs from the perspective of social choice and connect OWAs
to majority rules.
In the field of artificial intelligence, Goldsmith et al.~\cite{DBLP:conf/aaai/GoldsmithLMP14} generalized OWAs
as {\em rank-dependent scoring rules}.
For more details on OWAs, please refer to recent surveys by Yu et al.~\cite{DBLP:journals/air/YuPXY23} and by Csisz\'{a}r~\cite{9409788}.

Facility location games, which are also 
known as strategy-proof social choice with single-peaked preferences, has traditionally been studied in the literature of mechanism design and social choice~\cite{moulin:PC:1980}, and recently in the field of algorithmic game theory from the viewpoint of {\em approximate mechanism design}~\cite{procaccia:TEAC:2013}.
Various extensions have been proposed,
such as locating a facility on graph metric~\cite{schummer:JET:2004,dokow:EC:2012},
achieving better approximation ratios~\cite{lu:wine:2009,lu:EC:2010}
considering false-name manipulations (also known as Sybil attacks)~\cite{todo:AAMAS:2011,TodoOY:ECAI:2020,NehamaTY:JAAMAS:2022},
extending to multi-dimensional Euclidean spaces~\cite{sui:IJCAI:2013},
considering dynamic arrivals and departures~\cite{KeijzerW:IJCAI:2018,wada:AAMAS:2018}, and
locating multiple facilities~\cite{miyagawa:SCW:2001,serafino:ECAI:2014,sonoda:AAAI:2016,fong:AAAI:2018}.

Given difficulties of designing strategy-proof mechanisms in various mechanism design problems, obvious manipulability analysis~\cite{li:AER:2017,TROYAN2020104970} is a recent trend in the literature.
Ortega and Klein~\cite{ORTEGA2023515} proposed a two-sided matching mechanism that violates strategy-proofness but satisfies the non-obvious manipulability condition and analyze its performance over two famous algorithms, namely the deferred acceptance~\cite{Gale:AMM:1962} and the top-trading-cycles~\cite{shapley:JME:1974}.
Aziz and Lam also considered the same property for social choice settings, while they neither focused on facility location games nor ordered weighted average mechanisms~\cite{DBLP:conf/aldt/AzizL21}.

Considering fairness is also important in
social choice settings, including the facility location games. As we mentioned, the dictatorship mechanism is a powerful mechanism
that is always strategy-proof, but it is totally unfair in the sense that only the dictator agent's opinion is taken into account. 
Indeed, various fairness properties have been considered in facility location games, such as group-fairness~\cite{DBLP:conf/aaai/LiLC24,DBLP:conf/ijcai/ZhouLC22} and
egalitarian mechanisms~\cite{procaccia:TEAC:2013,LI2023113930}.
Wang et al.~\cite{DBLP:conf/aaai/0001WLC21} also considered fairness from the perspective of facilities. 
\section{Model}
\label{sec:model}

\begin{figure}[tb]
 \centering
 \begin{tikzpicture}[scale=1.0]
  \useasboundingbox [gray, dotted] (-1, -0.5) grid (7, 0.5);
  \draw [thick, draw = black] (0, 0) -- (6,0);
  \node at ([yshift = -10pt] 0, 0) {0};
  \node at ([yshift = -10pt] 6, 0) {1};
  \node [style=vote] at (1, 0) [label = {-90:\small{$x_{1}$}}] {};
  \node [style=vote] at (4, 0) [label = {-90:\small{$x_{2}$}}] {};
  \node [style=vote] at (5, 0) [label = {-90:\small{$x_{3}$}}] {};
  \node [style=outcome] at (2.5, 0) [label = 
  {-90:\small{$y = f(x)$}}] {};
 \end{tikzpicture}
 \caption{An example of facility location games. Agents' reported locations are represented as 
 circles. A mechanism determines where to locate a facility, represented as a square.}
 \label{fig:ex}
\end{figure}
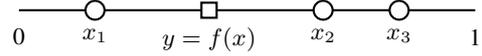

Let $A \subseteq \mathbb{R}$ be a closed interval in $\mathbb{R}$. In this paper, we assume $A := [0,1]$, but our model can be straightforwardly extended to an interval with an arbitrary length, say $[0, L]$.
Let $N$ denote the set of $n$ agents.
An agent $i \in N$ is assigned
a value $x_{i} \in A$, which is referred as
the agent $i$'s ideal location (or 
{\em peak}).
What we would like to achieve 
for a facility location game is 
a method, called a {\em mechanism},
which determines a location $y \in A$ of the facility by taking into account the reported
locations of those $n$ agents, which is illustrated in Fig.~\ref{fig:ex}. 

We then define the utility of agents.
Given two points $a, b \in A$,
$d(a,b) := |a-b|$ is the distance between $a$ and $b$.
Given a location $y \in A$ of the facility,
the {\em cost} of the agent $i$ with ideal location $x_{i}$ is given as the distance $d(x_{i}, y)$.
We then define the agent's {\em utility}
for given location $y$ and her ideal location $x_{i}$, 
for technical reasons, as one minus her cost,
that is,
$u(x_{i}, y) := 1 - d(x_{i}, y)$,
where the first term corresponds to the length 
of the outcome space.
Note that each agent's utility, as a function of the location $y$, is uniquely determined by the ideal location $x_{i}$. That is, the domain of preferences considered in this paper 
is a subclass of well-known {\em single-peaked preferences}~\cite{black:JPE:1948}. Indeed,
we assume mechanisms satisfy a property called {\em peaks-onliness}~\cite{EHLERS2002408}; a mechanism satisfies peaks-onliness if its output only depends on the peaks of the reported preferences.

Now we are ready to give the formal definition of decision making mechanisms and their desirable properties.
A deterministic decision making mechanism
(or just a mechanism in short) $f$
is defined as a mapping 
$f: A^{n} \rightarrow A$.
That is, it takes $n$ locations, which is usually denoted as a profile $x := (x_{i})_{i \in N}$, reported by the $n$ agents as an input and returns a point $f(x) \in A$.
For notation simplicity,
let $x_{-i}$ be a profile of $n-1$ locations
reported by agents except for $i$.
Let $f(x_{i}, x_{-i}) \in A$ indicate the location returned by mechanism $f$ when agent $i$ reports $x_{i}$ and the other agents
jointly reports $x_{-i} \in A^{n-1}$.

\begin{definition}[Pareto Efficiency]
 Given input $x = (x_{i})_{i \in N}$,
 a location $y \in A$ of a facility is said to 
 be {\em Pareto efficient} for $x$
 if there does not exist any other location $z \in A$ such that
 $u(x_{i}, z) \geq u(x_{i}, y)$ holds for every $i \in N$
 and
 $u(x_{i'}, z) > u(x_{i'}, y)$ holds for at least one $i' \in N$.
 A mechanism is said to be {\em Pareto efficient} (or satisfy PE)
 if 
 $f(x)$ is Pareto efficient 
 for any $x = (x_{i})_{i \in N} \in A^{n}$.
\end{definition}

It is clear by definition, and therefore well-known in the literature, 
that,
when agents' preferences are assumed to be
single-peaked,
a deterministic mechanism $f$ satisfies 
PE if and only if 
$ \min_{i \in N} x_{i} 
 \leq f(x)
 \leq \max_{i \in N} x_{i}$
holds for any input $x$.

\begin{definition}[Anonymity]
    A mechanism is said to be {\em anonymous}
    if, 
    for any $x = (x_{i})_{i \in N} \in A^{n}$,
    $f(x)
    = f(x_{\sigma})$
    holds, 
    where 
    $\sigma: N \rightarrow N$ is an arbitrary permutation
    and $x_{\sigma}$ be the permuted profile.
\end{definition}

Intuitively, an anonymous mechanisms 
treat all the agents equally, in the sense
that their names do not affect the outcome at all.

In the literature of social choice theory and facility location games,
several families of deterministic mechanisms have been proposed.
Generalized median voter schemes (GMVS) are 
a quite famous family of mechanisms, 
which are known to be the only mechanisms
simultaneously satisfying PE 
and strategy-proofness.
Furthermore, their anonymous subset, named as anonymous GMVS (AGMVS) in this paper, is the most well-investigated class of mechanisms in the literature. Moulin~\cite{moulin:PC:1980} showed that
AGMVS are the only mechanisms satisfying PE,
anonymity, and strategy-proofness.

\begin{definition}[Generalized Median Voter Schemes~\cite{moulin:PC:1980}]
    A mechanism $f$ is 
    a {\em generalized median voter scheme (GMVS)} if,
    there are $2^{n}-1$ parameters $(\alpha_{S})_{S \subseteq N} \in A^{2^{n}-1}$, each of which corresponds to a non-empty subset $S \subseteq N$,
    s.t.\ 
    for any input $x \in A^{n}$,
    $f(x) = \min_{S \subseteq N} \max_{i \in S} \{x_{i}, \alpha_{S}\}$.
\end{definition}

\begin{definition}[Anonymous GMVS~\cite{moulin:PC:1980}]
    A mechanism $f$ is 
    an {\em anonymous generalized median voter scheme (AGMVS)} if 
    there are $n-1$ parameters, $\beta_{1}, \beta_{2}, \ldots, \beta_{n-1}$ such that
    for any input $x \in A^{n}$,
    \[
        f(x) = \text{med}(x_{1}, x_{2}, \ldots, x_{n}, \beta_{1}, \beta_{2}, \ldots, \beta_{n-1}),
    \]
    where we assume the operator $\text{med}(\cdots)$ 
    returns the left-median, i.e., 
    the $n/2$-th smallest point,
    among any given set of points.
\end{definition}

The ordered weighted average (OWA) is another class of deterministic mechanisms, which is originally proposed by Yager~\cite{Yager:SMC:1988} for aggregation of multiple values.

\begin{definition}[Ordered Weighted Average~\cite{Yager:SMC:1988}]
    A mechanism $f$ is called an {\em ordered weighted average (OWA)} if there exists $n$ parameters $w_{1}, w_{2}, \ldots, w_{n}$, satisfying both $w_{j} \in [0,1]$ for any $j \in \{1, \ldots, n\}$ and $\sum_{j} w_{j} = 1$, such that for any input $x \in A^{n}$,
    \[
    f(x) = \sum_{j = 1}^{n} w_{j} x_{\pi(j)}
    \]
    where $\pi: N \rightarrow N$ is a permutation
    s.t.\ $x_{\pi(1)} \leq x_{\pi(2)} \leq \cdots \leq x_{\pi(n)}$~\footnote{In the literature, the permuted vector is usually represented as an descending order, i.e., $x_{\pi(1)} \geq x_{\pi(2)} \geq \cdots \geq x_{\pi(n)}$.
    In this paper we use an ascending order to make it consistent with the locations on the interval $A$.}.
\end{definition}

That is, an OWA has a normalized set of $n$ weights $(w_{j})_{j \in \{1, \ldots, n\}}$
and takes the weighted average for the {\em sorted} input $x_{\pi}$.
For example,
the center mechanism has the weights $w_{1} = w_{n} = 1/2$ and chooses the average
point between the minimum and the maximum reported locations, which is known to minimize the maximum cost objective~\cite{procaccia:TEAC:2013}.
The standard average mechanism has the
weights $w_{j} = 1/n$ for every $j$,
and takes the average of all the reported locations.
The Olympic average mechanism~\cite{DBLP:conf/aaai/GoldsmithLMP14}, also known as the trimmed mean or the truncated mean, has the weights $w_{1} = w_{n} = 0$ and $w_{j} = 1/(n-2)$ for every $j \neq 1, n$, which is used
in the Olympic games for judging, e.g., figure skating.
Also, order statistic mechanisms, which has the weight $w_{j} = 1$ for exactly one $j$, includes the median mechanism by choosing 
$j = \lceil \frac{n}{2} \rceil$.



\subsection{Incentive Properties}
\label{ssec:incentive}

Incentive properties have been one of the main interests in 
mechanism design,
where their objective is to incentivize agents
to behave sincerely; more specifically, in a direct revelation mechanism, we would like to incentivize agents to honestly report their private information, also called {\em type}, which corresponds to the ideal location in our model.
Among several incentive properties, strategy-proofness is one of the most well-studied properties, 
which requires that reporting a true type is a dominant strategy for every agent~\footnote{Please refer to, e.g., Nisan~\cite{Nisan_2007} for its formal definition.}. 

\begin{definition}[Strategy-Proofness]
    A mechanism satisfies {\em strategy-proofness (SP)}
    if, 
    for any $i \in N$,
    for any $x_{-i} \in A^{n-1}$,
    for any $x_{i} \in A$,
    and
    for any $x'_{i} \in A$,
    it holds that
    \begin{equation}
    \label{eq:SP}
    u(x_{i}, f(x_{i}, x_{-i}))
    \geq
    u(x_{i}, f(x'_{i}, x_{-i})).
    \end{equation} 
\end{definition}

Non-obvious manipulability (NOM) is a weakened property of strategy-proofness.
Instead of requiring that every manipulation is not beneficial,
NOM requires that in both the best and the worst case, according to the actions of the other agents, truth-telling is better than any other manipulation.

\begin{definition}[Non-Obvious Manipulability (NOM)]
 A mechanism is said to be {\em not obviously manipulable}
 (or satisfies {\em NOM}) 
 if,
 for any $i \in N$,
 for any $x_{i} \in A$,
 and
 for any $x'_{i} \in A$,
 both of the following inequalities hold:
 \begin{equation}
  \label{eq:NOM:best}
  \max_{x_{-i}}[u(x_{i}, f(x_{i}, x_{-i}))]
  \geq
  \max_{x_{-i}}[u(x_{i}, f(x'_{i}, x_{-i}))]        
 \end{equation}
 \begin{equation}
  \label{eq:NOM:worst}
  \min_{x_{-i}}[u(x_{i}, f(x_{i}, x_{-i}))]
  \geq
  \min_{x_{-i}}[u(x_{i}, f(x'_{i}, x_{-i}))]
 \end{equation}
\end{definition}

 If mechanism $f$ satisfies Eq.~\ref{eq:NOM:best} 
 (Eq.~\ref{eq:NOM:worst}, respectively),
 $f$ is said to satisfy NOM-B (NOM-W, resp.).
 By definition, a mechanism $f$ satisfies NOM
 if and only if it satisfies both NOM-B and NOM-W.
 By definition, SP implies NOM.

\subsection{Fairness Properties}
\label{ssec:fairness}
In the recent study of facility location games,
considering fairness among agents is a popular approach.
Proportionality-based fairness properties are known to be
a series of fairness benchmarks for analyzing social choice functions.
Here, we define five proportionality-based 
fairness properties, namely,
individual fair share (IFS),
unanimous fair share (UFS),
proportionally fairness (PF),
proportionality (P), and unanimity (UN).
Note that while the first three properties were
formally proposed by Aziz et al.~\cite{aziz2023strategyproof} by using the cost as a measure,
most of our discussions in this paper are based on the utilities. So we describe both their original
requirements based on distances and identical 
definitions based on utilities.

\begin{definition}[Individual Fair Share]
 \label{def:IFS}
 Given profile $x = (x_{i})_{i \in N}$ of reported locations,
 a location $y \in A$ of a facility is said to satisfy
 {\em individual fair share (IFS)} for $x$
 if
 \[
 d(x_{i}, y) \leq 1 - \frac{1}{n}
 \ \Big(
 \Leftrightarrow u(x_{i}, y) \ge \frac{1}{n}
 \ 
 \Big)
 \]
 holds for every $i \in N$.
 A mechanism $f$ is said to satisfy {\em individual fair share (IFS)} 
 if 
 $f(x)$ satisfies IFS 
 for any input $x = (x_{i})_{i \in N}$.
\end{definition}

\begin{definition}[Unanimous Fair Share]
 \label{def:UFS}
 Given profile $x = (x_{i})_{i \in N}$ of reported locations,
 a location $y \in A$ of a facility is said to satisfy
 {\em unanimous fair share (UFS)} for $x$
 if, for any coalition $S \subseteq N$
 such that $x_{j} = x_{j'}$ holds for some constant $x_{j'} = c \in A$ and every $j, j' \in S$, 
 \[
 d(x_{i}, y) \leq 1 - \frac{|S|}{n}
  \ 
  \Big(
  \Leftrightarrow u(x_{i}, y) \ge \frac{|S|}{n}
  \ 
  \Big)
 \]
 holds for every $i \in S$.
 A mechanism $f$ is said to satisfy {\em unanimous fair share (UFS)}
 if 
 $f(x)$ satisfies UFS 
 for any input $x = (x_{i})_{i \in N}$.
\end{definition}

\begin{definition}[Proportional Fairness]
 \label{def:PF}
 Given profile $x = (x_{i})_{i \in N}$ of reported locations,
 a location $y \in A$ of a facility is said to satisfy
 {\em proportional fairness (PF)} for $x$
 if, for any coalition $S \subseteq N$,
 \[
 d(x_{i}, y) \leq 1 - \frac{|S|}{n} + r
  \ 
  \Big(
  \Leftrightarrow u(x_{i}, y) \ge \frac{|S|}{n} - r
  \ 
  \Big)
 \]
 holds for every $i \in S$, where $r := \max_{j \in S} x_{j} - \min_{j \in S}x_{j}$.
 A mechanism $f$ is said to satisfy {\em proportional fairness (PF)} 
 if 
 $f(x)$ satisfies PF 
 for any input $x = (x_{i})_{i \in N}$.
\end{definition}

\begin{definition}[Proportionality]
 \label{def:proportionality}
 A mechanism $f$ is said to satisfy {\em proportionality (P)} 
 if, for any profile $x := (x_{i})_{i \in N} \in A^{n}$ such that $x_{i} \in \{0, 1\}$
 for all $i \in N$,
 it holds that
 \[
 f(x) = \frac{\# \{ i \in N \mid x_{i} = 1\}}{n}.
 \]
\end{definition}

\begin{definition}[Unanimity]
 \label{def:unanimity}
 A mechanism $f$ is said to satisfy {\em unanimity (UN)} 
 if, for any profile $x := (x_{i})_{i \in N} \in A^{n}$ satisfying $x_{i} = x_{i'}$ for every $i, i' \in N$ and some fixed $x_{i'}=c \in A$,
 $f(x) = c$ holds.
\end{definition}

Aziz et al.~\cite{aziz2023strategyproof}
investigated the relations among these fairness
properties and compatibility with strategy-proofness, which is summarized in the
following two claims.

\begin{theorem}[Aziz et al.~\cite{aziz2023strategyproof}]
 \label{thm:aziz-lam}
 A mechanism satisfies PF and SP
 if and only if it is a uniform phantom mechanism, i.e., the AGMVS
 whose $n-1$ parameters are set as
 $\beta_{l} = l/n$ for each $l \in \{1, \ldots, n-1\}$. 
\end{theorem}

\begin{proposition}[Aziz et al.~\cite{aziz2023strategyproof}]
 \label{prp:PF-UFS-IFS}
 PF implies UFS. UFS implies
 P, 
 IFS, and UN.
\end{proposition}


\section{Warm-Up: Strategy-Proof OWAs}
\label{sec:SP}
\begin{figure}[tb]
 \centering
 \begin{tikzpicture}[scale=1.0]
  \newcommand{\yone}{0}
  \newcommand{\ytwo}{-1.5}
  \useasboundingbox [gray, dotted] (-4, -2) grid (4, 0.5);
  %
  \draw [thick, draw = black] (-3, \yone) -- (3, \yone);
  \node at ([yshift = -10pt] -3, \yone) {0};
  \node at ([yshift = -10pt] 3, \yone) {1};
  \node [style=vote] at (-3, \yone) [label = {90:\small{$1, \ldots, j-1$}}] {};
  \node [style=vote] at (0, \yone) [label = {90:\small{$i$}}, label={270:\small{$x_{i} = \frac{1}{2}$}}
  ] {};
  \node [style=vote] at (3, \yone) [label = {90:\small{$j+1, \ldots, n$}}] {};
  \node [style=outcome] at (2, \yone) [label = {-90:\small{$f(x)$}}] {};
  \draw [dotted, black] (0, \yone) -- (0, \ytwo);
  \draw [thick, draw = black] (-3, \ytwo) -- (3, \ytwo);
  \node at ([yshift = -10pt] -3, \ytwo) {0};
  \node at ([yshift = -10pt] 3, \ytwo) {1};
  \node [style=vote] at (-3, \ytwo) [label = {90:\small{$1, \ldots, j-1$}}] {};
  \node [style=vote] at (-1, \ytwo) [label = {90:\small{$i$}}, label={270:\small{$x'_{i} = \frac{1}{2} - \epsilon$}}
  ] {};
  \node [style=vote] at (3, \ytwo) [label = {90:\small{$j+1, \ldots, n$}}] {};
  \node [style=outcome] at (1.6, \ytwo) [label = {-90:\small{$f(x'_{i}, x_{-i})$}}] {};
 \end{tikzpicture}
 \caption{A beneficial manipulation in OWA (except for order statistics), presented in the proof of Proposition~\ref{prp:SP}.}
 \label{fig:ex:SP}
\end{figure}
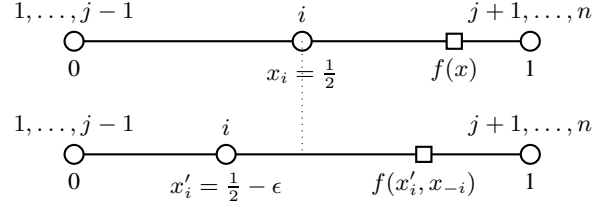

Since most of the existing works on facility location games focus on strategy-proof mechanisms, in this paper, we begin with analyzing the strategy-proofness (SP) of OWAs.
In the literature of approximate mechanism design
for facility location games,
it is well-known that the center mechanism,
which chooses the average of the minimum and the maximum locations in the input and thus minimizes the cost of the agent who has the highest cost, violates SP.
This means that not all OWAs satisfy SP,
but almost all the other OWAs have not been 
investigated in detail from the perspective of 
mechanism design.

Now, we show a necessary and sufficient condition
for OWAs to satisfy SP. Informally speaking,
an OWA satisfies SP if and only if 
it is an order statistic.

\begin{proposition}
 \label{prp:SP}
 An OWA satisfies SP if and only if
 $w_{j} = 1$ holds for some $j \in \{1, \ldots, n\}$.
\end{proposition}

\begin{proof}
 We first show the if direction.
 It is known that any OWA is Pareto efficient and anonymous~\cite{GARCIALAPRESTA2001463}.
 Therefore, from the characterization of strategy-proof,
 Pareto efficient, and anonymous mechanism by Moulin~\cite{moulin:PC:1980},
 it suffices to show that an OWA is represented as an AGMVS 
 if $w_{j} = 1$ holds for some $j \in \{1, \ldots, n\}$.
 Indeed, any given OWA $f$ that satisfies $w_{j} = 1$ for exactly one
 index $j \in \{1, \ldots, n\}$ is represented
 as an AGMVS whose parameters $(\beta_{1}, \ldots, \beta_{n-1})$
 are set so that $\beta_{k} = 0$ for any $k \in \{1, \ldots, n-j\}$
 and $\beta_{k} = 1$ otherwise, which returns the $j$-th order statistics.

 For the only-if direction,
 consider an arbitrarily chosen OWA $f$ in which
 at least two weights 
 are neither zero nor one. Let $j$ be
 the minimum such index satisfying $w_{j} \in (0,1)$.

 Let us consider a profile in which
 $j-1$ agents have a peak at $0$,
 one agent, say agent $i$, has a peak $x_{i} = \frac{1}{2}$, and
 the other $n-j$ agents have a peak on $1$.
 The OWA returns $w_{j} \cdot x_{i} + (1 - w_{j}) = 1 - \frac{w_{j}}{2}$ as the outcome
 when all the agents report their locations truthfully (see the top figure in Fig.~\ref{fig:ex:SP}).

 Now consider the case that 
 agent $i$ reports a different location $x'_{i} = \frac{1}{2} - \epsilon$,
 with a small positive real number $\epsilon$
 satisfying $\epsilon \leq \frac{1-w_{j}}{2w_{j}}$.
 The outcome then changes to 
 \[
 1 - \frac{w_{j}}{2} - w_{j}\epsilon
 \geq \frac{1}{2},
 \]
 which is strictly closer to the agent $i$'s true peak $x_{i} = \frac{1}{2}$ than
 the original outcome (see the bottom figure in Fig.~\ref{fig:ex:SP}). Therefore, the agent has an incentive 
 to misreport the location, violating strategy-proofness.
\end{proof}

This theorem implies that,
combined with the characterization theorem by Moulin~\cite{moulin:PC:1980},
the intersection between OWA and GMVS is represented
as the set of all order statistics.
That is, the following corollary holds,
which can be considered as a characterization of
the set of order statistics methods.
Note that since any OWA is anonymous,
we choose GMVS instead of AGMVS in the statement in order to avoid the duplication of anonymity property.

\begin{corollary}
 \label{crl:OWA-GMVS}
 A mechanism $f$ is represented as an order statistics
 if and only if $f$ is both an OWA and a GMVS.
\end{corollary}


\section{Not-Obviously-Manipulable OWAs}
\label{sec:NOM}

The necessary and sufficient condition in the previous section shows us quite a negative implication; strategy-proof OWAs must be somewhat ``unfair,'' in the sense that only one reported location always determines the location, 
and all the others may receive lower utilities.
We now consider weakening the incentive property
to non-obvious manipulability (NOM) and clarify
which OWAs satisfy that property.

We first show a general property of OWAs; any OWA satisfies NOM-B, regardless of their weights. Intuitively, as shown in the proof,
each agent can receive the best possible utility
{\em in the best case}. 

\begin{proposition}
 \label{prp:BNOM}
 Any OWA satisfies NOM-B.
\end{proposition}

\begin{proof}
 Let $f$ be an arbitrarily chosen OWA,
 and arbitrarily choose agent $i \in N$ and her true location $x_{i} \in A$.
 Since it is known that any OWA satisfies Pareto efficiency, 
 $f(x_{i}, x_{-i}) = x_{i}$ holds when
 every element in $x_{-i}$ coincides $x_{i}$.
 Then, it clearly holds that
 \[
 \max_{x_{-i}} [u(x_{i}, f(x_{i}, x_{-i})] = 1.
 \]
 Note that the LHS corresponds to the LHS of
 Eq.~\ref{eq:NOM:best}, and the value one in the RHS is the largest possible utility.
 Thus, this equation implies that Eq.~\ref{eq:NOM:best} always holds.
\end{proof}

Now we show one of our main results in this paper. An OWA satisfies NOM
if and only if either (i) it is an order statistic, or 
(ii) it assigns a zero weight to both the minimum and maximum reports.

\begin{theorem}
 \label{thm:NOM}
 An OWA satisfies NOM if and only if
 either (i) $w_{j} = 1$ for some $j \in \{1, \ldots, n\}$
    or
    (ii) $w_{1} = w_{n} = 0$.
\end{theorem}

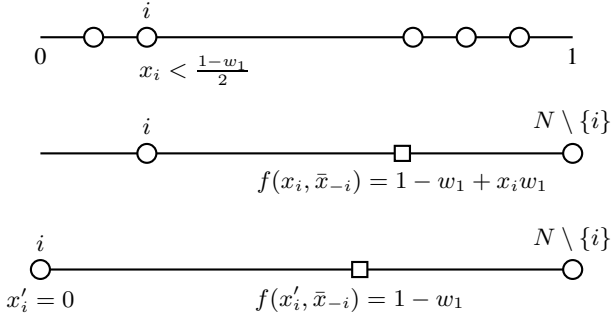
\begin{figure}[tb]
 \centering
 \begin{tikzpicture}[scale=.7]
  \newcommand{\yone}{0}
  \newcommand{\ytwo}{-2.2}
  \newcommand{\ythree}{-4.4}
  \useasboundingbox [gray, dotted] (-1, \ythree - 1) grid (11, \yone + 1);
  %
  \draw [thick, draw = black] (0, \yone) -- (10, \yone);
  \node at ([yshift = -10pt] 0, \yone) {0};
  \node at ([yshift = -10pt] 10, \yone) {1};
  \node [style=vote] at (2, \yone) [label = {90:\small{$i$}}, label = {-90:\small{$~~~~~~~~~~~~~~~~ x_{i}< \frac{1-w_{1}}{2}$}}] {};
  \node [style=vote] at (1, \yone) {};  
  \node [style=vote] at (7, \yone) {};  
  \node [style=vote] at (8, \yone) {};  
  \node [style=vote] at (9, \yone) {};  
  \draw [thick, draw = black] (0, \ytwo) -- (10, \ytwo);
  %
  \node [style=vote] at (2, \ytwo) [label = {90:\small{$i$}},
  ] {};  
  \node [style=vote] at (10, \ytwo) [label = {90:\small{$N \setminus \{i\}$}}] {};
  \node [style=outcome] at (6.8, \ytwo)
  [label = {-90:\small{$f(x_{i}, \bar{x}_{-i})
  = 1 - w_{1} + x_{i}w_{1}$}}] {};
  \draw [thick, draw = black] (0, \ythree) -- (10, \ythree);
  \node [style=vote] at (0, \ythree) [label = {90:\small{$i$}}, label = {-90:\small{$x'_{i} = 0$}}] {};  
  \node [style=vote] at (10, \ythree) [label = {90:\small{$N \setminus \{i\}$}}] {};
  \node [style=outcome] at (6, \ythree)
  [label = {-90:\small{$f(x'_{i}, \bar{x}_{-i}) = 1 - w_{1}$}}] {};
 \end{tikzpicture}
 \caption{The profiles used in the proof of the 
 only if part of Theorem~\ref{thm:NOM}. The top 
 indicates the original input, the middle 
 indicates the worst-case for truth telling $x_{i}$, and 
 the bottom indicates the worst-case for
 manipulation $x'_{i} = 0$. E.g., parameters $w_{1} = 0.4$ and $x_{i} = 0.2$ works for this 
 example.}
 \label{fig:proof:NOM}
\end{figure}

\begin{proof}
 Proving the if direction for condition (i), which is used in
 Proposition~\ref{prp:SP}, is obvious from the fact 
 that SP implies NOM.
 Combined with Proposition~\ref{prp:BNOM},
 it then suffices to show that 
 OWAs satisfy NOM-W under condition (ii).
 
 Let $R_f(x_{i})$ be the set of possible outcomes of a given mechanism $f$, under the condition that
 agent $i$ reports $x_{i}$. Formally,
 $R_{f}(x_{i}) :=  \{ y \in A \mid \exists x_{-i}, y = f(x_{i}, x_{-i}) \}$.
 When $f$ is an OWA satisfying condition (ii),
 the two extreme positions among all the reported locations
 are assigned the weight zero, and thus have no effect on the outcome.
 It is then obvious that for any $x_{i} \in A$,
 $R_{f}(x_{i}) = A$; for any $y \in A$,
 we can choose a profile $x_{-i} = (x_{i'})_{i' \neq i}$ such that
 $x_{i'} = y$ holds for any $i' \neq i$, which returns 
 $f(x_{i}, x_{-i}) = y$.
 Therefore, $R_{f}(x_{i}) = R_{f}(x'_{i})$ holds for any pair $x_{i}, x'_{i}$ of locations, which implies the following target inequality, namely,
 \[
 \min_{x_{-i}} [u(x_{i}, f(x_{i}, x_{-i}))]
 \geq
 \min_{x_{-i}} [u(x_{i}, f(x'_{i}, x_{-i}))].
 \]
 Indeed, the worst case utility is given as $\min (x_{i}, 1 - x_{i})$ for both sides of the inequality.

 We then prove the only if direction. For the sake of
 contradiction, we assume that both conditions (i) and (ii)
 are violated in the weights of a given OWA mechanism $f$. We then show that $f$ violates NOM. The input profiles used in this direction
 are summarized in Fig.~\ref{fig:proof:NOM}.

 The weights of an OWA violate both conditions (i) and (ii) simultaneously
 if and only if either $w_{1} \in (0,1)$ or $w_{n} \in (0,1)$ holds.
 From symmetry, we assume, without loss of generality, that
 $w_{1} \in (0,1)$ holds.
 Let us then consider the case where a manipulating agent $i$
 has a true location $0 < x_{i} < \frac{1-w_{1}}{2}$ (see the top figure in Fig.~\ref{fig:proof:NOM}). 

 It is known by Yager~\cite{Yager:SMC:1988} that any OWA mechanism is monotonic. Therefore, the worst possible outcome for $x_{i}$ is given under the profile either (i) $\bar{x}_{-i} := (x_{i'})_{i' \neq i}$ such that $x_{i'} = 1$ holds for any $i' \in N \setminus \{i\}$ or (ii) $\bar{x}_{-i} := (x_{i'})_{i' \neq i}$ such that $x_{i'} = 0$ holds for any $i' \in N \setminus \{i\}$
 regardless of the report $x'_{i}$.
 When agent $i$ reports $x_{i}$, 
 the outcome for (i) is $1 - w_{1} + x_{i}w_{1}$ and that for (ii) is $x_{i}w_{n}$.
 Thus, the utility for (i) is $w_{1} - x_{i}w_{1} + x_{i}$ and that for (ii) is 
 $1 - x_{i} + x_{i} w_{n}$.
 From the assumption that $x_{i} < \frac{1-w_{1}}{2}$ we have 
 $w_{1} - x_{i}w_{1} + x_{i} < 1 - x_i + x_{i}w_{n}$
 and thus
 the minimum utility is $w_{1} - x_{i}w_{1} + x_{i}$,
 which corresponds to the LHS of Eq.~\ref{eq:NOM:worst}, i.e., the inequality for NOM-W.
 On the other hand, the outcome when agent $i$ reports $x'_{i} = 0$ is $1 - w_{1} (\geq x_{i})$ for (i) and $0$ for (ii), and the respective utility is $w_{1} + x_{i}$ and $1 - x_{i}$.
 Again, from the assumption that $x_{i} < \frac{1-w_{1}}{2}$ we have $w_{1} + x_{i} < 1 - x_{i}$ and 
 therefore the minimum utility is 
 $w_{1} + x_{i}$, which corresponds to the RHS of Eq.~\ref{eq:NOM:worst}.
 Since both $w_{1}$ and $x_{i}$ are positive numbers, 
 the LHS is strictly smaller than the RHS, which violates NOM (more specifically, NOM-W).
 Actually, as shown in Fig.~\ref{fig:proof:NOM},
 the facility gets closer (from the middle to the bottom) to the manipulating agent $i$'s true location $x_{i}$.
\end{proof}

The assumption on the true location, $x_{i} < \frac{1-w_{1}}{2}$, is introduced to 
guarantee that the worst case for the manipulator $i$
occurs when all the others $i' \neq i$ report 
$x_{i'} = 1$, which makes the proof simpler.
If we choose $w_{1} = 0.4$ and $x_{i} = 0.4$,
the assumption is violated, and indeed,
the worst case for the misreport $x'_{i} = 0$
is when all the others report $0$, 
while the worst case for the truth-telling
$x_{i}$ is when all the others report $1$.
Indeed, the misreport gives her a higher utility $0.4$ than the worst case utility $0.36$ under her truth-telling. 

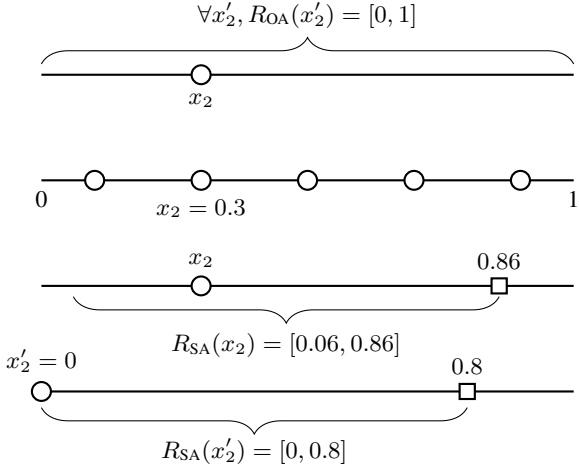
\begin{figure}[tb]
 \centering
 \begin{tikzpicture}[scale=.7]
  \newcommand{\yone}{0}
  \newcommand{\ytwo}{-2}
  \newcommand{\ythree}{-4}
  \newcommand{\yfour}{-6}
  \useasboundingbox [gray, dotted] (-1, \yfour - 1.5) grid (11, \yone + 1.5);
  %
  \draw [thick, draw = black] (0, \yone) -- (10, \yone);
  %
  \draw [decorate,decoration={brace,amplitude=10pt,raise=1ex}]
  (0,\yone) -- (10,\yone) node[midway,yshift=2.5em]{$\forall x'_{2}, R_{\text{OA}}(x'_{2}) = [0,1]$};  
  \node [style=vote] at (3, \yone) [label = {-90:\small{${x_{2}}$}}] {};
  \draw [thick, draw = black] (0, \ytwo) -- (10, \ytwo);
  \node at ([yshift = -10pt] 0, \ytwo) {0};
  \node at ([yshift = -10pt] 10, \ytwo) {1};
  \node [style=vote] at (1, \ytwo) [] {};
  \node [style=vote] at (3, \ytwo) [label = {-90:\small{${x_{2}} = 0.3$}}] {};
  \node [style=vote] at (5, \ytwo) [] {};
  \node [style=vote] at (7, \ytwo) [] {};
  \node [style=vote] at (9, \ytwo) [] {};
    %
  \draw [thick, draw = black] (0, \ythree) -- (10, \ythree);
   \draw [decorate,decoration={brace,amplitude=10pt,mirror,raise=1ex}]
  (0.6,\ythree) -- (8.6,\ythree) node[midway,yshift=-2.5em]{$R_{\text{SA}}(x_{2}) = [0.06, 0.86]$}; 
  \node [style=vote] at (3, \ythree) [label = {90:\small{${x_{2}}$}}] {};
  \node [style=outcome] at (8.6, \ythree)
  [label = {90:\small{$0.86$}}] {};
  \draw [thick, draw = black] (0, \yfour) -- (10, \yfour);
   \draw [decorate,decoration={brace,amplitude=10pt,mirror,raise=1.5ex}]
  (0, \yfour) -- (8,\yfour) node[midway,yshift=-2.5em]{$R_{\text{SA}}(x'_{2}) = [0, 0.8]$}; 
  \node [style=vote] at (0, \yfour) [label = {90:\small{${x'_{2} = 0}$}}] {};
  \node [style=outcome] at (8, \yfour)
  [label = {90:\small{$0.8$}}] {};
  \end{tikzpicture}
 \caption{Example~\ref{ex:OA-SA} illustrates the 
 key difference between the Olympic average (OA) and the standard average (SA). 
 The second top figure shows a true input.
 The top figure shows that the Olympic average 
 satisfies NOM, while the bottom two figures show that the standard average violates NOM-W.}
 \label{fig:ex:AVG}
\end{figure}

To explain the intuition of the theorem,
let us show the following example where both
the average mechanism and the Olympic average 
mechanism are applied. More specifically,
we illustrate the effect of setting the weights of an OWA as $w_{1} = w_{n} = 0$ in Fig.~\ref{fig:ex:AVG}.
Note that both of these mechanisms are not strategy-proof.

\begin{example}
 \label{ex:OA-SA}
 Assume there are five agents, $N = \{1, \ldots, 5\}$,
 whose true locations are given as
 $x_{i} = 0.2 \cdot i - 0.1$ (see the second top figure in Fig.~\ref{fig:ex:AVG}). 

 First, let us consider the case where the Olympic average applies,
 i.e., an OWA mechanism whose weights are set as
 $w_{1} = w_{5} = 0$, and $w_{j} = \frac{1}{3}$ 
 for $j = \{2,3,4\}$. Note that it is 
 not strategy-proof;
 agent $2$ with true location $x_{2} = 0.3$ has 
 an incentive to report, say, $x'_{2} = 0$,
 which changes the outcome from $0.5$ to 
 $0.433...$, violating Eq.~\ref{eq:SP}.

 However, such a manipulation is not {\em obviously}
 beneficial in the worst case for agent 2. 
 When she tells the truth, the worst possible case for her true type is that all the other agents report $1$, which gives her the weight zero, and thus, the outcome is $1$. In this case, her utility,
 i.e., the LHS of Eq.~\ref{eq:NOM:worst} is $1 - (1 - x_{2}) = 0.3$. 
 Even when she tells $x'_{2} = 0$, the worst possible case for her true type, not for her misreport, is still the same, so that all the other agents report $1$.
 Therefore, her utility, i.e., the RHS of Eq.~\ref{eq:NOM:worst} is still $1 - (1 - x_{2}) = 0.3$. Comparing these two cases, the NOM-W condition holds. 

 We then consider that the standard average applies,
 i.e., an OWA mechanism whose weights are set as 
 $w_{j} = \frac{1}{5}$ for every $j \in N$, 
 and show that the above manipulation 
 is obviously beneficial in the worst case.
 When agent 2 reports her ideal location truthfully, the worst case is when all the other agents report $1$, in which the outcome is $0.86$ (see the second bottom figure in Fig.~\ref{fig:ex:AVG}). Thus, the LHS of Eq.~\ref{eq:NOM:worst} 
 is $1 - (0.86 - x_{2}) = 0.44$.
 When agent 2 reports $x'_{2} = 0$, the worst 
 case is still in the same case where all the other 
 agents report $1$, in which the outcome is $0.8$ (see the bottom figure in Fig.~\ref{fig:ex:AVG}). 
 Thus, the RHS of Eq.~\ref{eq:NOM:worst} 
 is $1 - (0.8 - x_{2}) = 0.5$, which is strictly 
 greater than the LHS, violating the NOM-W condition.
\end{example}

The key difference between the Olympic average and the standard average can be explained as follows. 
In the Olympic average, any agent is assigned the weight zero at the worst case, where she has no effect on the outcome. Therefore, $R_{\text{OA}}(x'_{i}) = [0,1]$
holds for any $x'_{i}$ (as shown in the top figure in Fig.~\ref{fig:ex:AVG}), meaning that 
all the possible outcomes in $A$ are realizable
according to the reports $x_{-i}$ of the other agents, regardless of the report $x'_{i}$ by agent $i$.
In contrast, in the standard average, every agent {\em always} has a strictly positive (and uniform) weight and thus has some effect on the worst case outcome. Indeed, the set $R_{\text{SA}}(x'_{i})$
of realizable outcomes varies when we choose a different $x'_{i}$, as shown in the bottom two figures in Fig.~\ref{fig:ex:AVG}.

\section{Proportionality-Based Fairness}
\label{sec:fairness}

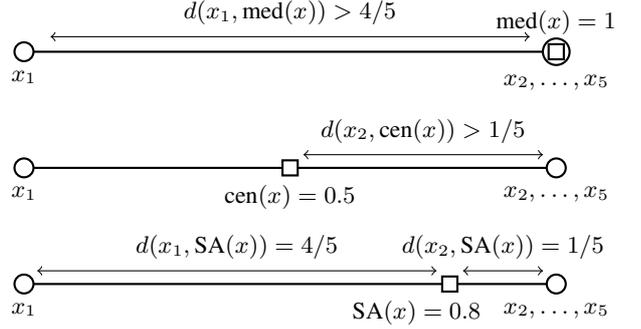
\begin{figure}[tb]
 \centering
 \begin{tikzpicture}[scale=.7]
  \newcommand{\yone}{0}
  \newcommand{\ytwo}{-2.2}
  \newcommand{\ythree}{-4.4}
  \useasboundingbox [gray, dotted] (-1, \ythree - 1) grid (11, \yone + 1.5);
  %
  \draw [thick, draw = black] (0, \yone) -- (10, \yone);
  \node [style = vote] at (0, \yone) [label = {-90:\small{${x_{1}}$}}] {};
  \node [style = vote, minimum size = 10pt] at (10, \yone) [label = 
  {-90:\small{${x_{2}, \ldots, x_{5}}$}}] {};
  \node [style = outcome] at (10, \yone) 
  [label = {90: ${\text{med}(x) = 1}$}]{};
  \draw [<->] (0.5, \yone+2ex) -- (9.5, \yone+2ex) node[midway,yshift=1em]{$d(x_{1}, \text{med}(x)) > 4/5$};
  %
  %
  %
  \draw [thick, draw = black] (0, \ytwo) -- (10, \ytwo);
  \node [style = vote] at (0, \ytwo) [label = {-90:\small{${x_{1}}$}}] {};
  \node [style = vote] at (10, \ytwo) [label = 
  {-90:\small{${x_{2}, \ldots, x_{5}}$}}] {};
  \node [style = outcome] at (5, \ytwo)
  [label = {-90: ${\text{cen}(x) = 0.5}$}]{};
  \draw [<->] (5.25, \ytwo + 0.25) -- (9.75, \ytwo + 0.25) node[midway,yshift=1em]{$d(x_{2}, \text{cen}(x)) > 1/5$};
 %
  \draw [thick, draw = black] (0, \ythree) -- (10, \ythree);
  \node [style = vote] at (0, \ythree) [label = {-90:\small{${x_{1}}$}}] {};
  \node [style = vote] at (10, \ythree) [label = 
  {-90:\small{${x_{2}, \ldots, x_{5}}$}}] {};
  \node [style = outcome] at (8, \ythree)  [label = {-90: ${\text{SA}(x) = 0.8~~~~~~~~~~~}$}]{};
  \draw [<->] (8.25, \ythree + 0.25) -- (9.75, \ythree + 0.25) node[midway,yshift=1em]{$d(x_{2}, \text{SA}(x)) = 1/5$};
  \draw [<->] (0.25, \ythree + 0.25) -- (7.75, \ythree + 0.25) node[midway,yshift=1em]{$d(x_{1}, \text{SA}(x)) = 4/5$};
 \end{tikzpicture}
 \caption{Agent 1 is at 0 and other four agents 2, ..., 5 are at 1. The top figure shows that the median mechanism violates IFS, the middle 
 figure shows that the center mechanism violates PF, and the bottom figure shows that the 
 standard average mechanism satisfies PF.}
 \label{fig:PF}
\end{figure}

We now turn to discuss which proportionality-based fairness properties can be satisfied by OWA mechanisms.
Note that unanimity is known to be implied by the Pareto efficiency property.
Since any OWA satisfies Pareto
efficiency, we have the following observation.

\begin{observation}
 \label{obs:OWA:UN}
 Any OWA satisfies unanimity.
\end{observation}

We first investigate IFS, i.e., for every agent
and every possible situation,
her cost is required to be bounded by $1 - \frac{1}{n}$.
We give a necessary and sufficient condition
on the weights of OWAs to guarantee IFS as follows.

\begin{theorem}
 \label{thm:IFS}
 Any OWA mechanism satisfies IFS if and only if 
 \[
  w_{1} \geq \frac{1}{n}
  \text{ \  and \  }
  w_{n} \geq \frac{1}{n}.
 \]
\end{theorem}

\begin{proof}
 We first show the if direction.
 In any OWA, given an arbitrary ideal location $x_{i}$ of an agent $i$, her cost is maximized when all the other agents are located at the farthest away point from $x_{i}$, which is either 0 or 1. 
 Without loss of generality, assume that $x_{i} \geq \frac{1}{2}$.
 The other case is analogous.
 In this case, $x_{i}$ is the largest reported location, and all the others are located at 0, therefore the OWA places the facility at $w_{n}x_{i}$, which gives her the cost $(1-w_{n}) x_{i}$.
 Since $w_{n} \geq 1/n$ is assumed and
 $x_{i} \leq 1$ holds in our model,
 $(1-w_{n}) x_{i} \leq 1 - \frac{1}{n}$
 holds. The LHS corresponds to her cost,
 and thus, this inequality coincides with 
 the definition of IFS.
 
 For the only if part, let us assume without loss of generality that $w_{n} < \frac{1}{n}$ (the case $w_{1} < \frac{1}{n}$ is analogous).
 Then, let us consider the following input $x$:
 \[
 x := (\underbrace{0, \ldots, 0}_{n-1}, 1).
 \]
 From the definition of OWA, it locates the facility
 at $w_{n} < \frac{1}{n}$, in which the 
 agent located at 1 has a cost strictly larger than
 $1 - \frac{1}{n}$, violating IFS. 
\end{proof}

Intuitively, OWAs must give higher priorities (i.e., larger weights) to both the maximum and minimum extreme locations to achieve IFS. An example of such an OWA is the center mechanism, which has weights $w_{1} = w_{n} = 1/2$ and, therefore, chooses the average of the maximum and minimum extreme locations.

We also give a necessary and sufficient condition for the weights to guarantee PF,
a stronger property than IFS.
Indeed, as the following theorem states,
the OWA mechanism satisfying PF is uniquely determined, which is the standard average mechanism.

\begin{figure}[tb]
 \centering
 \begin{tikzpicture}[scale=1.0]
  \useasboundingbox [gray, dotted] (-1, -0.25) grid (7, 2.25);
  \node (org-PF) at (1, 0) {PF};
  \node (org-UFS) at (1, 1) {UFS};
  \node (org-P) at (0, 2) {P};
  \node (org-IFS) at (1, 2) {IFS};
  \node (org-UN) at (2, 2) {UN};
  \draw[->] (org-PF) -- (org-UFS);
  \draw[->] (org-UFS) -- (org-P);
  \draw[->] (org-UFS) -- (org-IFS);
  \draw[->] (org-UFS) -- (org-UN);
  \node (new-PF) at (5, 0) {PF = UFS = P};
  \node (new-IFS) at (5, 1) {IFS};
  \node (new-UN) at (5, 2) {UN};
  \draw[->] (new-PF) -- (new-IFS);
  \draw[->] (new-IFS) -- (new-UN);
  \end{tikzpicture}
 \caption{Relations among proportionality-based 
 fairness properties,
 where arrows represent implications. The left 
 figure is from Aziz et al.~\cite{aziz2023strategyproof}. The right 
 figure shows 
 their relations by focusing only on OWAs.}
 \label{fig:summary:fairness}
\end{figure}
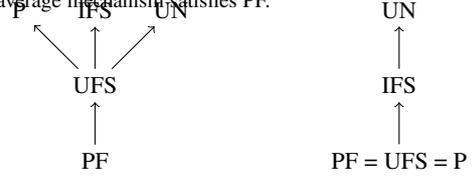

\begin{theorem}
 \label{thm:PF}
 An OWA mechanism satisfies PF
 if and only if 
 \begin{equation}
 \label{eq:OWA:PF}
 \forall j \in \{1, \ldots, n\}, \quad w_{j} = \frac{1}{n}.
 \end{equation}
\end{theorem}

\begin{proof}
 For the if direction, we show that the 
 standard average mechanism $f$ satisfies PF.
 For the sake of contradiction,
 assume that there is a coalition $S \subseteq N$ and an agent $i \in S$
 such that, for some input $x$,
 \[
  d(x_{i}, f(x)) > 1 - \frac{|S|}{n} + r
 \]
 holds, where $r := \max_{j \in S} x_{j} - \min_{j \in S} x_{j}$.
 Now let $p = \min_{j \in S} x_{j}$
 and $q = \max_{j \in S} x_{j}$.
 Obviously, $q = p + r$ holds.
 
 In that case, for at least one extreme agent $i'$ located at $x_{i'} \in \{p, q\}$, which is possibly equal to $i$, 
 it holds that 
 \[
 d(x_{i'}, f(x)) \geq d(x_{i}, f(x)) 
 >1 - \frac{|S|}{n} + r.
 \]
 In other words, either 
 $x_{i} \in [p, f(x)]$
 or
 $x_{i} \in [f(x), q]$ holds, 
 since $x_{i} \in [p, q]$ holds by definition. 
 Here we assume $x_{i'} = p$
 and thus $x_{i} \in [p, f(x)]$;
 a similar argument applies for $x_{i'} = q$.

 Now consider another input $x'$, by modifying $x$, where all the agents in $S$ except $i'$ (if any) increase their locations to $q$. From the monotonicity of OWA $f$,
 $f(x) \leq f(x')$
 holds.
 We then modify input $x'$ and obtain another input $x''$,
 where all the agents $N \setminus S$ (if any) increase their locations to $1$.
 From monotonicity again, 
$f(x') \leq f(x'')$
 holds.
 Since 
 the coalition $S$ has the same range $r$
 in both $x$ and $x''$,
 we have
 \begin{equation}
 \label{eq:PF}
 d(x_{i'}, f(x'')) > 1 - \frac{|S|}{n} + r.     
 \end{equation}

 Note that, in the input $x''$,
 there are one report at $p$,
 $|S| - 1$ reports at $q = p + r$,
 and $n - |S|$ reports at $1$.
 Thus, $f(x'')$ is determined as
 \[
 f(x'') = 
 \frac{|S|}{n} p
 + \frac{|S| - 1}{n} r
 + \frac{n-|S|}{n},
 \]
 and therefore,
 \[
 d(p, f(x'')) = 
 \frac{|S| - n}{n} p
 + \frac{|S| - 1}{n} r
 + \frac{n-|S|}{n}
 \leq 1 - \frac{|S|}{n} + r
 \]
 holds, which violates Eq.~\ref{eq:PF}.

 For the only if direction,
 we show that any OWA whose weights violate the above condition fails to satisfy the proportionality (Def.~\ref{def:proportionality}),
 i.e., even violates a weaker notion of proportionality-based fairness.
 
 Assume Eq.~\ref{eq:OWA:PF} is violated.
 We can then find a value $j \in \{1, \ldots, n-1\}$ such that 
 $\sum_{j' = 1}^{j} w_{j'} \neq 
 j/n
 $.
 Now let us consider the input profile $x$ such that
 \[
 x := (\underbrace{0, \ldots, 0}_{j}, \underbrace{1, \ldots, 1}_{n-j}).
 \]
 For this input $x$, the OWA returns 
 $1 - \sum_{j' = 1}^{j} w_{j'} \neq \frac{n-j}{n}$
 as an outcome, which
 violates the proportionality constraint.
\end{proof}

Observation~\ref{obs:OWA:UN}
and Theorems~\ref{thm:IFS} and \ref{thm:PF} jointly highlight the difference among
the median, the center, and the standard average mechanisms.
The median satisfies unanimity as well as SP.
The standard average satisfies PF, while
the center only satisfies IFS,
while these two mechanisms violate even NOM.
The intuition of the reason why the center violates PF is that it ignores all but the two extreme locations, which are considered in the standard average. For example, see Fig.~\ref{fig:PF}; there is one agent at point zero and the other four agents at point one,
which is represented as an input $x$.
The median returns $\text{med}(x) = 1$ as the outcome (see the top figure), violating IFS for agent $1$; $d(x_{1}, \text{med}(x)) > 1 - \frac{1}{5}$.
The center returns $\text{cen}(x) = 0.5$ as the outcome (see the middle figure), which satisfies IFS but violates PF 
for subset $S = \{2,3,4,5\}$;
$d(x_{2}, \text{cent}(x)) > 1 - \frac{4}{5}$. The standard average, shown at the bottom, satisfies PF.

From those findings presented in this section, 
we have the following 
corollary and observation, which is
summarized in the right figure of Fig.~\ref{fig:summary:fairness}.
Note that the left figure in Fig.~\ref{fig:summary:fairness}
is originally drawn in
Aziz et al.~\cite{aziz2023strategyproof} for general facility location games.

\begin{corollary}[from Theorem~\ref{thm:PF}]
 \label{crl:OWA-PF-proportionality}
 Any OWA satisfying P 
 also satisfies PF (and therefore, UFS).
\end{corollary}

\begin{observation}
 \label{obs:OWA:summary}
 Focusing on OWAs, P 
 (or, equally, PF or UFS) implies IFS,
 and IFS implies UN.
\end{observation}





\section{Discussions and Concluding Remarks}
\label{sec:conclu}

We have observed several necessary and
sufficient conditions for two distinct 
criteria, namely incentive and fairness.
What we consider here is 
when these two criteria go together.
Table~\ref{tbl:summary} in the beginning summarizes their (in)compatibility,
given our theoretical findings. 
Indeed, as a corollary of those results, the following holds.

\begin{corollary}
 \label{crl:OWA-NOM-IFS}
 There is no OWA mechanism that simultaneously 
 satisfies NOM and IFS.
 On the other hand, any OWA mechanism
 satisfies both NOM-B and UN.
\end{corollary}

This indicates a quite sharp
incompatibility between the incentive and proportionality-based fairness; while the minimum requirement for each of those criteria, i.e., NOM-B and UN, is always achievable, stronger properties in the two criteria cannot be achieved together.
For example, the Olympic average mechanism, which is shown to satisfy NOM in this paper,
clearly violate proportionality (and thus, PF) 
since it totally ignores the two extreme agents.

On the other hand, if we consider the anonymity 
property also as a fairness requirement,
in the sense that all the agents should be treated equally,
it can automatically be achieved by any OWA by definition.
Given these observations, defining another class
of fairness properties and showing their compatibility with incentive properties
would be a promising future direction.

About the model of facility location games, although our model can be easily extended
to any closed interval $A \subseteq \mathbb{R}$
with an arbitrary length $L$,
our analysis of both incentive and fairness strictly depends on the assumption that
agents' cost is defined {\em exactly by the distance}. This means that, considering
general single-peaked preferences, as
studied in e.g., Moulin~\cite{moulin:PC:1980},
is not straightforward.
On the other hand, applying some extensions of OWAs to facility location games might be possible, such as focusing only on some set of discrete values in the interval~\cite{DBLP:journals/isci/Llamazares07}.

As other potential directions,
providing a complete characterization of 
a class of mechanisms, e.g., the OWA mechanisms,
possibly by using some incentive properties such as NOM-B would also be theoretically interesting,
as Moulin~\cite{moulin:PC:1980} did with strategy-proofness and Todo et al.~\cite{todo:AAMAS:2011} did with
false-name-proofness.
It might also be possible to analyze
under which input agents have the incentive to misreport in OWAs; as the easiest case,
in the center mechanism, no agent has an incentive to misreport when the minimum and the maximum agents are located at zero and one, respectively.



\begin{ack}
We would like to thank anonymous reviewers for their valuable comments. 
This work was partially supported by JST ERATO Grant Number JPMJER2301, and JSPS
KAKENHI Grant Numbers JP21H04979 and JP20H00587, Japan.
\end{ack}

\end{document}